\documentclass[letter,reprint,amsmath,amssymb,aps,twocolumn,nofootinbib,superscriptaddress]{revtex4-1}

\usepackage{graphicx}
\usepackage{dcolumn}
\usepackage{bm}
\usepackage[hidelinks,
colorlinks=True,
citecolor=magenta,
urlcolor=red,
linkcolor=blue]{hyperref}
\usepackage{cleveref}
\usepackage{float}
\usepackage[normalem]{ulem}
\usepackage[dvipsnames]{xcolor}
\newcommand\be{\begin{equation}}
\newcommand\ee{\end{equation}}

\newcommand\ph{\phantom{s}}
\newcommand\dd{{\rm d}}

\newcommand\mL{\mathcal}

\begin{document}

\title{Multifield curved solid: Early dark energy and perturbation instabilities}

\author{Juan P. Beltrán Almeida}
\email{jubeltrana@unal.edu.co}
\affiliation{Universidad Nacional de Colombia, Facultad de Ciencias, Departamento de Física,\\
Avenida Cra 30 \# 45-03, Bogotá, Colombia}

\author{Alejandro Guarnizo}
\email{aguarnizo50@uan.edu.co}
\affiliation{Departamento de F\'isica, Universidad Antonio Nari\~no, \\ Cra 3 Este \# 47A-15, Bogot\'a, Colombia}

\author{Thiago S. Pereira}
\email{tspereira@uel.br}
\affiliation{Departamento de F\'isica, Universidade Estadual de Londrina, \\
Rodovia Celso Garcia Cid, Km 380, 86057-970, Londrina, Paran\'a, Brazil}

\author{C\'esar A. Valenzuela-Toledo}
\email{cesar.valenzuela@correounivalle.edu.co}
\affiliation{Departamento de F\'isica, Universidad del Valle,\\
Ciudad Universitaria Mel\'endez, Santiago de Cali 760032, Colombia}

\date{\today}

\begin{abstract}
We introduce a multifield dark energy model with a nonflat field-space metric, in which one field is dynamical while the others have constant spatial gradients. The model is predictive at the background level, leading to an early dark energy component at high redshifts and a suppressed fraction of late-time anisotropy. Both features have simple expressions in terms of the curvature scale of the field-space, and correspond to stable points in the phase space of possible solutions. Because of the coupling between time and space-dependent scalar fields, vector field perturbations develop tachyonic instabilities at scales below the Hubble radius, thus being potentially observable in the number count of galaxies. Overall, the presence of a nontrivial field-space curvature also leads to the appearance of instabilities on scalar perturbations, which can impact the matter density distribution at large scales.
\end{abstract}

\maketitle

\section{\label{sec:1}Introduction}

The existence of an unexplained dark sector continues to be one of the main drivers of new ideas in cosmology, forcing us to test the limits of the standard lambda cold dark matter ($\Lambda\text{CDM}$) model. Recently this program has gained momentum due to the realization that measurements of critical cosmological parameters, among which $H_0$, give different values when done using high or low redshift data~\cite{Abdalla:2022yfr,Aluri:2022hzs}. Alongside these facts, there remain numerous large-angle cosmic microwave background (CMB) anomalies which also lack an explanation---fundamental or not---and that contribute to the suspicion that the $\Lambda\text{CDM}$ model is just a first approximation to a more accurate description of the universe.

Among the many potential candidates to explain these problems lies the possibility that dark energy results from new matter degrees of freedom. The most popular and paradigmatic example is the quintessence field and its many avatars~\cite{Tsujikawa:2013fta}.\footnote{Though it has been argued that low $z$ data favors quintessence models leading to lower $H_0$ in comparison to $\Lambda\text{CDM}$ --- see~\cite{Banerjee:2020xcn}} At first sight, the versatility of quintessence models seems to be limited by the imposition of translation invariance arising from the cosmological principle, which forces the fields to be homogeneous. However, a closer inspection of the Lagrangian from the simplest quintessence model,
\[
{\mathcal L} = -\frac{1}{2}\partial_\mu\phi\partial^\mu\phi\ - V(\phi)\,,
\]
reveals that, as far as the kinetic term is concerned, translation invariance does not require homogeneity of the \emph{fields}, but rather the homogeneity of their spatial gradients. The realization of this possibility has led to many interesting implementations known generically as ``solid'' or ``elastic'' models, where the dynamics is achieved by means of \emph{space-dependent} scalar fields with constant spatial gradients. To account for a forbidden (symmetry breaking) potential, such models usually rely on a noncanonical Lagrangian capable of reproducing an accelerated expansion. This idea has its roots on
effective field-theoretic implementations of solid and elastic media~\cite{Dubovsky:2011sj,Ballesteros:2016gwc}, and has since found fertile ground in cosmological applications, being implemented in inflation~\cite{Gruzinov:2004ty,Endlich:2012pz,Bartolo:2013msa}, dark matter~\cite{Bucher:1998mh,Verlinde:2016toy}, and dark energy~\cite{Armendariz-Picon:2007umg,Motoa-Manzano:2020mwe}.

In the standard applications of solids to cosmology, the internal symmetries of the fields, such as SO(3) and shift invariances, are explored when building phenomenologically viable cosmological models~\cite{Endlich:2012pz,Celoria:2017bbh}. In this regard, one could also envisage the existence of a curvature in field-space, as is usually done in nonlinear sigma models of quantum field theory. In fact, \emph{time-dependent} nonlinear sigma models have been long explored in the context of inflation~\cite{Sasaki:1995aw,Langlois:2008mn,Allen:2005ye,Achucarro:2010jv, Achucarro:2010da,Kaiser:2012ak,Renaux-Petel:2015mga} and dark energy \cite{Vardanyan:2015oha,Akrami:2020zfz,Cicoli:2020cfj,Anguelova:2021jxu,Eskilt:2022zky}. Recently, the authors of Ref.~\cite{Akrami:2020zfz} considered the implementation of dark energy in one such model, showing that a phase of accelerated expansion can be achieved even with steep potentials, provided that trajectories in field-space are sufficiently non-geodesic.

In this work we ask whether a time-dependent field could coexist with a solid, thus bridging a gap between the solid and nonlinear multifield descriptions. Specifically, we introduce a nonlinear dark energy model in which one field is dynamical, while three other have constant spatial gradients. The model is predictive at the background level, and can lead to a percent level fraction of early dark energy, as well as a suppressed fraction of late-time anisotropy. We show that the early dark energy component is proportional to the square of the field-space curvature radius, while late-time anisotropies has an upper bound given by the fourth
power of this radius. Thus, for small and observationally interesting radii, the anisotropy is always suppressed compared to early dark energy. When small scale linear perturbations are included, we show that the existence of a non-trivial curvature in field space leads to instabilities in transverse vector modes, with a characteristic length scale of order of the present Hubble radius. Such instabilities will feed gravitational vector perturbations, which could, in principle, be observed through the abundance of large-scale structures. The same curvature term is responsible for the appearance of gravitational instabilities in scalar perturbations, though the derivation of a dispersion relation and the length scales of these instabilities is hard to obtain analytically. Overall, our findings corroborate the claim that a negative field-space Ricci scalar can lead to gravitational instabilities~\cite{Renaux-Petel:2015mga}. 

The mathematical formalism appropriate for our description, known as harmonic maps, is quickly reviewed in Sec. \ref{sec:harmonic-maps}. In Sec. \ref{sec:curved-solid} we discuss the main idea of the curved solid model, and derive its main features at the background level. The dynamics of small scale linear instabilities is analyzed in Sec. \ref{sec:perturbations}, after which we conclude and give some perspectives of future developments in Sec. \ref{sec:conclusion}. Throughout this work we adopt (modified) Planck units in which $\hbar=c=8\pi G=1=M_\text{Pl}^{-2}$ and a metric signature $(-,+,+,+)$.

\section{Harmonic Maps}\label{sec:harmonic-maps}

We start by giving a concise introduction to the formalism of harmonic maps, which underpins our implementation. Further details can be found in~\cite{eells1964harmonic,Misner:1978am,eells1978report}. Although the mathematics is the same as that used in multifield modeling (see, e.g., \cite{Sasaki:1995aw,Achucarro:2010jv,Renaux-Petel:2015mga}), many physically interesting insights and possible applications introduced by the formalism seem to be unknown to (most) cosmologists. We thus hope that this section can also help drawing attention to these important references. Readers interested in our implementation can skip to Sec. \ref{sec:curved-solid}.

The formalism of harmonic maps consists in a set of scalar fields $\phi^A$ evolving over a space-time ${\cal M}$ with metric $g_{\mu\nu}$, such that the fields themselves define a Riemannian space $\mL{N}$, the \emph{field-space}, with metric $G_{AB}$. The dimension of $\mL{N}$ is arbitrary, although for the applications we have in mind $\text{dim}\,\mL{N}=4$; the field-space indices $A, B, \dots$ thus run from 0 to 3. 

Because the fields live in the space-time, they map points from ${\cal M}$ to ${\cal N}$, since for any point $x^\mu$ in ${\cal M}$ there corresponds a point $\phi^A$ in ${\cal N}$ through the functions $\phi^A(x^\mu)$. This identification allows us to pull field-space objects back to the space-time through the matrix $\partial_\mu\phi^A$. This is possible because, under general field [$\phi^A \rightarrow \psi^A(\phi^C)$] and coordinate [$x^\mu \rightarrow y^\mu(x^\nu)$] transformations, this matrix transforms linearly:
\begin{equation}
\frac{\partial\psi^A}{\partial y^\mu}= \frac{\partial\psi^A}{\partial\phi^C}\frac{\partial x^\nu}{\partial y^\mu}\frac{\partial\phi^C}{\partial x^\nu}\,.
\end{equation}
Thus, for example, the field-space metric $G_{AB}$ and connection $\Gamma^A_{BC}$ can be pulled back from the field-space to the space-time as
\begin{align}
G_{\mu\nu} & \equiv \partial_\mu\phi^A\partial_\nu\phi^B G_{AB}\,, \\
\quad
\Gamma^A_{\mu B} & \equiv \partial_\mu\phi^C \Gamma^A_{CB}\,.
\end{align}
This property renders the dynamics of the fields nonlinear, and it serves as the central feature from which many of the interesting properties of multifield cosmological models emerge, as well as many of interesting properties of harmonic maps in general~\cite{Misner:1978am}.

Given this structure, we can take \emph{space-time} covariant derivatives of field-space tensors provided we attach, for each field-space index, a corresponding connection term $\Gamma^A_{\mu B}$. Thus, for example, the space-time covariant derivative of any vector $v^A$ in ${\cal N}$ will be
\begin{align}
\nabla_\mu v^A & = \partial_\mu v^A + \Gamma^A_{\mu B}v^B\,, \\
& = \partial_\mu\phi^B(\partial_B v^A+\Gamma^A_{BC}v^C)\,,\\
& = \partial_\mu\phi^B\nabla_B v^A\,.
\end{align}
For a tensor with mixed indices, $t^{AB\mu\nu\dots}$, one proceeds as usual,  adding the corresponding connection 
$\Gamma^\mu_{\lambda\nu}$ for each space-time index, and $\Gamma^A_{\mu B}$ for each field-space index. Note however that $\nabla_\mu\phi^A$ is still given by $\partial_\mu\phi^A$, as usual. From these definitions and the condition of metric compatibility in field space, $\nabla_A G_{BC}=0$, one readily verifies that 
\begin{equation}
\nabla_\mu G_{AB}=0\,. 
\end{equation}
This is an important property, and ensures that the manipulation of field-space indices commutes with space-time derivatives; in particular, it allows us to write $G_{AB}\nabla_\mu\phi^B$ simply as $\nabla_\mu\phi_A$.

The fields $\phi^A$ are called harmonic when they extremize the action 
\be\label{eq:harmonic-action}
S=-\frac{1}{2}\int\dd^4x\sqrt{|g|}\,\nabla^\mu\phi^A\nabla_\mu\phi_A\,.
\ee
Physically, if $\mL{M}$ describes a rubber and $\mL{N}$ a body, then $S$ gives the elastic energy needed to wrap the rubber around the body, while the fields obey an equation describing the configuration of zero elastic tension on the rubber~\cite{eells1978report}:
\be\label{eq:harmonic-eom}
\nabla^\mu\nabla_\mu\phi^A \equiv
\frac{1}{\sqrt{|g|}}\partial_\mu(\sqrt{|g|}\partial^\mu \phi^A) + \Gamma^A_{\mu B}\partial^\mu\phi^B 
 = 0 \,.
\ee
where $g=\text{det}(g_{\mu\nu})$. This equation makes evident the nonlinearity of the dynamics.

Known examples of harmonic maps include the Klein-Gordon equation for a massless scalar field 
(take $\text{dim}\,{\cal M}=4$ and $\text{dim}\,{\cal N}=1$) and the geodesic equation ($\text{dim}\,{\cal M}=1$, $\text{dim}\,{\cal N}=4$). A less obvious example is given by the geodesic deviation equation~\cite{Misner:1978am}. 

\section{Curved solid}\label{sec:curved-solid}

The vanilla harmonic model~\eqref{eq:harmonic-action} cannot lead to accelerating cosmologies. For that we also need a potential $V(\phi^A)$. We thus consider a generalized harmonic model where the fields are minimally coupled to gravity:
\begin{equation}\label{action}
S = \int\dd^4x\sqrt{|g|}\left[\frac{1}{2}R - \frac{1}{2}\nabla^\mu\phi^A\nabla_\mu\phi_A - V + {\cal L}_m\right],
\end{equation}
with ${\cal L}_m$ being the matter Lagrangian. Variation of this action then leads to the following nonlinear equation
\begin{equation}\label{eq:klein-gordon}
\nabla^\mu\nabla_\mu\phi^A -G^{AB}\partial_B V = 0\,.
\end{equation}

The class of models in which we are interested exhibit the most versatility in universes described by \emph{homogeneous} metrics of the 
following form:
\be\label{eq:spacetime-metric}
\dd s^2 = -\dd t^2 + a^2(t)\gamma_{ij}(t)\dd x^i \dd x^j\,,
\ee
with the condition that $\textrm{det}(\gamma_{ij})=1$ in order to preserve comoving volumes -- see~\cite{Pereira:2007yy} for details. These include the isotropic and spatially flat Friedmann-Lema\^itre-Robertson-Walker (FLRW) model (${\dot{\gamma}}_{ij}=0$), as well as the anisotropic Bianchi-I model (${\dot{\gamma}}_{ij}\neq0$). In the latter, $a(t)$ represents the mean scale factor, whereas the anisotropic expansion is characterized by the shear tensor
\begin{equation}\label{eq:sigma}
\sigma_{ij} \equiv \frac{1}{2}\dot{\gamma}_{ij}\,,
\end{equation}
where a dot means $\dd/\dd t$ and spatial indices are manipulated with $\gamma_{ij}$.
These two metrics can be encapsulated by Misner's parametrization, where one writes (no sum over $i$)
\begin{equation}\label{BetaBianchi}
 \gamma_{ij} = [e^{2\beta_i}]\delta_{ij}\,,\qquad \sum_i {\beta_i} = 0
\end{equation}
with the inverse given by $\gamma^{ij}=[e^{-2\beta_i}]\delta^{ij}$. In terms of $\beta_i$, the shear tensor becomes
\begin{equation}\label{def-beta}
 \sigma^{i}_{\ph j} = [\dot{\beta}_i]\delta^i_j\,.
\end{equation}
The isotropic case follows naturally by taking $\beta_i(t)=0$.

In this setup, the energy-momentum tensor of the fields acquires its most general form:
\begin{equation}\label{eq:emt}
T_{\mu\nu} = (\rho+p)u_\mu u_\nu + p g_{\mu\nu} + 2q_{(\mu}u_{\nu)} + \pi_{\mu\nu}\,,
\end{equation}
where the field's momentum density ($q^\mu$) and anisotropic stress ($\pi_{\mu\nu}$) satisfy the usual constraints: $q^\mu u_\mu = \pi^\mu_{\ph\mu} = 0 = u^\mu \pi_{\mu\nu}$. Here, $u^\mu$ is the four-velocity of fundamental observers, normalized as $u^\mu u_\mu=-1$. In the comoving coordinates defined by~\eqref{eq:spacetime-metric} it reads $u^\mu = (1,0,0,0)$. For these observers \eqref{eq:spacetime-metric}, the nonzero components of \eqref{eq:emt} are given by
\begin{align}
 \rho & = \frac{1}{2}\left[\dot{\phi}^A\dot{\phi}_A + \frac{1}{a^2}\partial^k\phi^A\partial_k\phi_A\right] + V(\phi^A) \,,\label{eq:rho-def} \\
 p & = \frac{1}{2}\left[\dot{\phi}^A\dot{\phi}_A -\frac{1}{3a^2}\partial^k\phi^A\partial_k\phi_A\right] - V(\phi^A)\,,\label{eq:p-def} \\
 q^i & = -\dot\phi^A \partial^i\phi_A\,,\label{eq:qi-def}\\
 \pi^i_{\ph j} & =  \frac{1}{a^2}\left(\partial^i\phi^A\partial_j\phi_A  - \frac{1}{3}\partial^k\phi^A\partial_k\phi_A\delta^i_j\right)\label{eq:pi-def}\,.
\end{align}

Fundamental observers are uniquely defined by the condition that the total momentum density vanishes in their rest frame~\cite{ellis2012relativistic}. For a model containing matter ($m$), radiation ($r$) and the fields $\phi^A$, this means that $q^i_m+q^i_r+q^i=0$. After recombination, matter and radiation only interact gravitationally, so that $q^i_m=0=q^i_{r}$. Thus, for our model to be compatible with the symmetries of \eqref{eq:spacetime-metric} at late times, we need \eqref{eq:qi-def} to be zero as well. Given that $q^i$ couples time and spatial derivatives, there are three distinct classes of models where it vanishes:
\begin{enumerate}
\item All fields are time-dependent: $\phi^A = \phi^A(t)$; 
\item All fields are space-dependent: $\phi^A = \phi^A(x^i)$;
\item A subset of fields are time-dependent, while the remaining are space-dependent. 
\end{enumerate}
Incidentally, each one of these configurations automatically leads to $\rho$, $p$ and $\pi^i_{\ph j}$ compatible with the symmetries of~\eqref{eq:spacetime-metric}. Let us analyze each case individually.

Models of class 1 consists of the usual multifield cosmological models that have been used to described either inflation and dark energy. In the context of dark energy, it was recently implemented in a two field model using flat~\cite{Akrami:2020zfz} and nonflat~\cite{Eskilt:2022zky} field-space metrics. Note that in this case not only $q^i$ but also $\pi_{ij}$ is zero. Since the latter behaves as a source to the shear, the trace-free part of the Einstein equation [see Eq.~\eqref{eq:trace-free}] implies that $\sigma^2 \sim a^{-6}$. Thus, assuming that the amplitude of the shear is consistent with CMB constraints, typically $(\sigma/H)_0\lesssim10^{-10}$~\cite{martinez1995delta,Saadeh:2016sak}, one can safely disregard the contribution of this term to the expansion at late times. In other words, harmonic models with time-dependent fields are (virtually) isotropic.

Models where the fields depend only on spatial coordinates (class 2) are commonly referred to as ``solid'' or ``elastic'' models~\cite{Bucher:1998mh,Gruzinov:2004ty,Endlich:2012pz}. Fields with an arbitrary spatial dependence cannot be generally employed, though, since they violate spatial homogeneity. However, because the energy-momentum tensor of the fields depends only on their gradients and a potential $V$, one \emph{can} build  cosmologically viable models if the fields are given by linear functions of position and if $V=0$~\footnote{One could also adopt the weaker condition of a constant $V$, but this is tantamount to a adding a cosmological constant to the model.}. For these models, homogeneity also requires that the field-space metric $G_{AB}$ be constant. As an example, consider the case of three fields defined as $\phi^I = x^I$, $I=(1,2,3)$, and a field-space with flat metric $\delta_{IJ}$. In this case $\nabla_I=\partial_I$ and the fields $\phi^I$ satisfy Eq.~\eqref{eq:klein-gordon} identically. Note however that, even though the fields have no dynamics, the matter sector of these models is nontrivial, since the stress tensor is nonzero
\begin{equation}
\pi^i_{\ph j} = \frac{1}{a^2}\left( \gamma^{ij} - \frac{1}{3}\gamma^{kk}\,\delta^i_j\right)\,.
\end{equation}
Moreover, the dynamics is also nontrivial, since \eqref{eq:rho-def} introduces an effective curvature term to the expansion rate: $\rho \sim a^{-2}$. Thus, for time-dependent $\gamma_{ij}$, the \emph{effective} background dynamics is that of a Bianchi-I universe with spatial curvature.

Before concluding our discussion on the models of class 2, it should be noted that space-dependent fields with nontrivial dynamics can be implemented by means of an effective field theory approach. In this case, one substitutes the kinetic term in~\eqref{action} with the most general function of SO(3) invariant operators, which are in turn constructed from products and contractions of the gradients $\partial_i\phi^I$ with itself and the metric $\delta_{IJ}$. This was explored both in the context of inflation \cite{Endlich:2012pz,Bartolo:2013msa}, dark matter \cite{Bucher:1998mh} and dark energy~\cite{Armendariz-Picon:2007umg,Motoa-Manzano:2020mwe}. At leading order in a derivative expansion, the first nontrivial operator is given by $g^{ij}\partial_i\phi^I\partial_j\phi^I \delta_{IJ}$. In view
of the formalism of harmonic maps, this is just the trace of the pushed-forward metric $g^{IJ}=g^{ij}\partial_i\phi^I\partial_j\phi^J$ with the flat field-space metric $G_{IJ}=\delta_{IJ}$. Thus, the formalism of harmonic maps provides a framework for nontrivial geometric generalizations of the standard (flat) solid models.

Class 3 models represent a novel possibility which, to the best of our knowledge, has not been explored~\footnote{In the context of flat field-space metrics, time and space-dependent fields were considered in \cite{Celoria:2017bbh}.}. Given the preceding discussion, we shall refer to them as \emph{curved solid} models. As we are going to show, when implemented as a dark energy component, they predict small anisotropies as well as an early dark energy phase. We now give the details of one concrete implementation.

\subsection{Concrete implementation}

For concreteness we are going to consider a four field model in which the first, $\phi^0$, is dynamical, while the other three, $\phi^I$, are pure spatial gradients. According to our previous discussion, the potential $V$ cannot depend on $\phi^I$. We thus define
\begin{equation}\label{def:solid-model0}
\phi^0\equiv\varphi(t), \qquad \phi^I \equiv x^I, \qquad V\equiv V(\varphi)\,.
\end{equation}
We shall informally refer to $\varphi$ and $\phi^I$ as quintessence and solid fields, respectively. 
As for the field-space metric $G_{AB}$, it can only be fixed by a more fundamental theory. However, similar to the potential $V$, cosmological applications require that $G_{AB}$ be solely a function of $\varphi$. Given that we can always choose synchronous coordinates in which $G_{00}=1$, a symmetric and nontrivial field-space metric is that of a flat and ``expanding'' field-space:
\begin{equation}\label{eq:fieldspace-metric}
\dd\ell^2 = \dd\varphi^2 + f^2(\varphi)\delta_{IJ}\dd\phi^I\dd\phi^J\,,
\end{equation}
where $f(\varphi)$ is an arbitrary function of $\varphi$. The nonzero Christoffel symbols for this metric are
\begin{equation}
\Gamma^0_{IJ} = -ff_{,\varphi}\,\delta_{IJ}\,,\qquad
\Gamma^I_{0J} = \frac{f_{,\varphi}}{f}\,\delta_{IJ}\,,
\end{equation}
where the ``comma $\varphi$'' notation means $\dd/\dd\varphi$. Using these expressions and the condition $\partial_I V = 0$, one readily verifies that the solids $\phi^I$ satisfy Eq.~\eqref{eq:klein-gordon}. The quintessence field, on the other hand, obeys a Klein-Gordon equation~\footnote{Note that $H = \dot{a}/a$, where $a$ is the mean scale factor defined in~\eqref{eq:spacetime-metric}.},
\begin{equation}
\ddot{\varphi} + 3H\dot{\varphi} + U_{,\varphi} = 0\,,
\end{equation}
where $U$ is a combination of the quintessence potential and a term coming from the kinetic coupling in \eqref{eq:klein-gordon}
\begin{equation}\label{Ueff}
U(\varphi,t) \equiv \frac{\gamma(t)}{2a^2}f^2(\varphi) + V(\varphi)\,,
\end{equation}
and where
\begin{equation}
\gamma(t)\equiv\textrm{tr}(\gamma^{ij})\,.
\end{equation}
This equation also follows from the continuity equation, $\dot{\rho}+3H(\rho+p)=-\sigma_{ij}\pi^{ij}$, with the energy density, pressure and anisotropic stress given by
\begin{align}
\rho_\varphi & = \frac{1}{2}\dot{\varphi}^2 + \frac{\gamma f^2}{2a^2} + V\label{def-rho_varphi}\,, \\
p_\varphi & = \frac{1}{2}\dot{\varphi}^2 - \frac{\gamma f^2}{6a^2} - V\label{def-p_varphi}\,, \\
\pi^i_{\ph j} & = \frac{f^2}{a^2}\left(\gamma^{ij}-\frac{1}{3}\gamma\delta^i_j\right)\label{eq:pi-model}\,,
\end{align}
respectively. It is interesting to note that while ${\rho_\varphi = \dot{\varphi}^2/2 + U}$,
as in typical quintessence models, ${p_\varphi \neq \dot{\varphi}^2/2 - U}$ (this was already anticipated by Eqs. \eqref{eq:rho-def} and \eqref{eq:p-def}). Such asymmetry follows from us having three spacelike vectors $\partial^\mu\phi^I$ and only one timelike vector $\partial^\mu\varphi$ contributing to the energy-momentum tensor (see also~\cite{Armendariz-Picon:2007umg}). Incidentally, this fact prevents an accelerated phase given exclusively by the field space metric, i.e., a phase where $f(\varphi)$ plays the role of a slow-roll potential. Indeed, by setting $V=0$ we find
\begin{equation}
1\geq p_\varphi/\rho_\varphi\geq -1/3\,,
\end{equation}
which, as we see, is not sufficiently negative to accelerate the expansion of the universe. 

The full background dynamics is given by the following equations
\begin{align}
3H^2 & = \rho_m + \rho_r + \rho_\varphi  + \frac{\sigma^2}{2}\,,\label{eq:H2}\\
(\sigma^i_{\ph j})^{\boldsymbol{\cdot}} & = - 3H\sigma^i_{\ph j} + \pi^i_{\ph j}\,,\label{eq:trace-free}\\
\ddot{\varphi} & = -3H\dot{\varphi} - \frac{\gamma}{a^2}ff_{,\varphi} - V_{,\varphi}\label{eq:KG-varphi}\,.
\end{align}
where $\sigma^2=\sigma_{ij}\sigma^{ij}$. In order to solve this system of equations we need to specify the potentials $V(\varphi)$ and $f(\varphi)$. For the purpose of illustrating the features of the model, it suffices to consider simple exponentials given by
\begin{equation}\label{def:solid-model1}
V(\varphi)=V_0 e^{-\lambda\varphi}\,,\quad f(\varphi)= f_0 e^{-\mu\varphi}\,,
\end{equation}
where $\lambda$ and $\mu$ measure the steepness of the potentials, while $V_0$ and $f_0$ measure their amplitudes. It is important to note that while $V_0$ is measured in units of $[\textrm{mass}]^4$, $f_0$ is dimensionless, since it is a component of the field-space metric. However, since the function $f(\varphi)$ always comes multiplied by combinations of $\partial^i\phi^I\partial_j\phi^J$, which is also measured in units of $[\textrm{mass}]^4$, the constant $f_0$ works in practice as a measure of the solid's energy scale~\footnote{To put it differently, we could have defined  $\phi^I = \sqrt{f_0} x^I$, and   $f(\varphi)=e^{-\mu\varphi}$, since what matters is the product $f(\varphi)\partial^i\phi^I\partial_j\phi^J$.}. Thus, in what follows, we will interpret $f_0$ as an (effective) energy scale. 

Given \eqref{def:solid-model1}, the field-space metric \eqref{eq:fieldspace-metric} represents a maximally symmetric manifold with Riemann tensor given by
\begin{equation}\label{Riemann}
\mL{R}_{ABCD} = \frac{\mL{R}}{4(4-1)}(G_{AC}G_{BD}-G_{AD}G_{BC})\,.
\end{equation}
where $\mL{R}$ is the associated Ricci scalar,
\begin{equation}\label{Ricci}
\mL{R} = -12\mu^2 \equiv -\frac{12}{(\ell_c)^2}\,,
\end{equation}
and where, in analogy to the de Sitter metric, we have introduced a curvature scale through $\ell_c = \mu^{-1}$. Thus, the parameter $\mu$ sets the curvature scale of the field-space. As we will see, the main aspects of the early dark energy stage (EDE), including its onset and amplitude, depend directly on
$\mu$ and $f_0$, and only indirectly on $\lambda$ and $V_0$, which in turn govern the late-time regime of dark energy. Our results are however fairly robust against the choice~\eqref{def:solid-model1} if compared with other viable cosmological potentials. Moreover, the choice we adopt here is more easily adapted to a dynamical system analysis, which we present in a separate work~\cite{BigPaper:2023}. For the numerical analysis it is also convenient to introduce the usual fractional densities:
\begin{equation}\label{def-Omega}
\Omega_\alpha \equiv \frac{\rho_\alpha}{3H^2}\,,\quad\alpha=(m,r,\varphi)\,,
\end{equation}
as well as the dark-energy equation of state: $w_\varphi = p_\varphi/\rho_\varphi$. Next, we present a concrete numerical implementation of this model and discuss some of its main features in a semianalytical approach.

\subsection{Early dark energy}\label{subsec:EDE}

\begin{figure}
\includegraphics[width=0.9\linewidth]{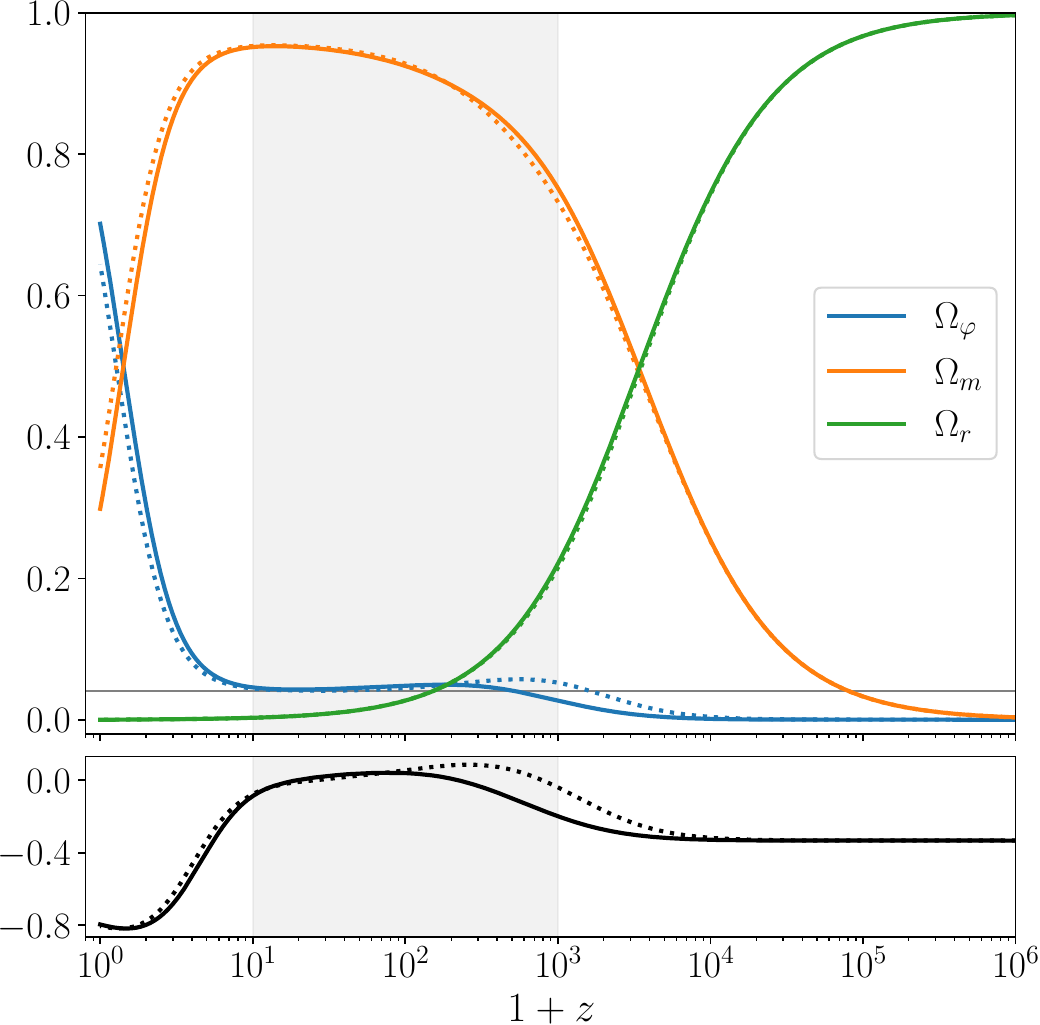}
\caption{Top: matter (orange), radiation (green), and dark-energy (blue) fractions as a function of the redshift. The thin horizontal line represents the approximate fraction of dark energy during the EDE phase (Eq.~\eqref{Omega_varphi-approx}), while the shaded region shows the approximate duration of this regime. Bottom: dark energy equation of state. For this plot we have adopted $\mu=2\lambda=2$, $f_0^2/H_0^2=2V_0/H_0^2=30$ (continuous lines) and $f_0^2/H_0^2=8V_0/H_0^2=120$ (dotted lines).}\label{fig:EarlyDeFLRW}
\end{figure}

Let us consider first the case of an isotropic FLRW universe, where $\beta_{i}=0$. Thus
\begin{equation}
\sigma^i_{\ph j}=0=\pi^i_{\ph j}\,.
\end{equation}
Since the matter sector is not coupled to the fields, $\rho_m$ and $\rho_r$ decays as $a^{-3}$ and $a^{-4}$, as usual, so that we only need to solve Eqs.~\eqref{eq:H2} and~\eqref{eq:KG-varphi}. A typical solution is shown in Fig.~\ref{fig:EarlyDeFLRW}, where we see an EDE regime starting at $z\simeq1000$ and contributing to about 5\% of the cosmological budget.

To understand the qualitative behavior of this figure, and in particular how the EDE phase is triggered, we can ignore the contribution of $V(\varphi)$ to the dynamics, since its contribution is only relevant at late times ($z\simeq0$). After the radiation era, and assuming EDE is subdominant, the universe is mostly matter dominated. If written in terms of the scale factor, Eq.~\eqref{eq:KG-varphi} becomes
\begin{equation}\label{eq:varphi-approx}
\frac{\dd ^2\varphi}{\dd a^2} + \frac{5}{2a}\frac{\dd\varphi}{\dd a} \approx \frac{\gamma\mu f_0^2e^{-2\mu\varphi}}{H_0^2\Omega_m^0a}\,.
\end{equation}
Note that while $\gamma=3$ in a FLRW universe, the expression above, and the ones following from it, hold for arbitrary $\gamma(t)$. We thus keep it as $\gamma$ for later reference. In what follows we shall consider that $\mu$ is a nonzero and positive constant, although our final results do not depend on its sign.

The field $\varphi$ starts at zero at the top of the potential $U$. During matter domination, while $\varphi$ is slowly rolling down, $2\mu\varphi\approx 0$, and the exponential in the equation above is frozen at 1. The solution satisfying ${\varphi(a=0)=0}$ is given by
\begin{equation}\label{eq:phi_onset}
\varphi = \frac{2}{5}\frac{\gamma\mu f_0^2}{H_0^2\Omega^0_m}a\,.
\end{equation}
This approximate solution, valid during matter domination, holds until $2\mu\varphi\lesssim 1$. When $2\mu\varphi\approx 1$ the exponential term, and hence $U$, decay faster. The field then gains kinetic energy, and EDE starts. This happens at a redshift $1+z_* = a_*^{-1}$ given by
\begin{equation}\label{zstar}
1+z_* \approx \frac{4}{5}\frac{\gamma\mu^2 f_0^2}{H_0^2\Omega^0_m}\,.
\end{equation}
For the parameters used in Fig.~\eqref{fig:EarlyDeFLRW} this gives $1+z_* \approx 10^3$, which is in excellent agreement with the (full) numerical solution shown in the same figure. As $\varphi$ continues to grow, the contribution of the potential $f(\varphi)$ becomes negligible in comparison to $V(\varphi)$, and the standard dark energy regime starts (this happens at $z\simeq 10$ in Fig.~\eqref{fig:EarlyDeFLRW}). The amplitude of $\Omega_\varphi$ during this regime, like the onset of EDE, is also controlled by the parameter
$\mu$. To see how this happens, note from Fig.~\eqref{fig:EarlyDeFLRW} that $\Omega_\varphi$ tracks the evolution of $\Omega_m$ during matter domination. Thus, $\dot\varphi^2 \sim a^{-3}$ or, in terms of the scale factor, $\dd\varphi/\dd a = c/a$, for some constant $c$. This constant can be fixed by demanding the continuity of $\dd \varphi/\dd a$ at $a_*$. From the derivative of \eqref{eq:phi_onset} and Eq.~\eqref{zstar} it follows that $c=1/(2\mu)$. To proceed, we substitute $\dd\varphi/\dd a = c/a$ and $\dd^2\varphi/\dd a^2 = -(1/a)\dd\varphi/\dd a$ into Eq.~\eqref{eq:varphi-approx}, which then gives
\begin{equation}\label{eq:approx_f}
\gamma f_0^2 e^{-2\mu\varphi} \approx  \frac{3}{4}\frac{H_0^2\Omega_m^0}{\mu^2a}\,.
\end{equation}
If we now use this result in~\eqref{def-rho_varphi}, still neglecting the potential $V$, it follows after some algebra that~\footnote{We stress that the limit $\mu=0$ cannot be taken in this expression, since it violates our previous assumption $2\mu\varphi=1$.}
\begin{equation}\label{Omega_varphi-approx}
\Omega_\varphi \approx \frac{1}{6\mu^2}\,,\qquad (10\lesssim z \lesssim 10^3)\,.
\end{equation}
Note that, in view of \eqref{Ricci}, $\Omega_\varphi$ is proportional to the square of the field-space curvature radius during EDE phase: $\Omega_\varphi\propto\ell_c^2$. For $\mu=2$, this expression gives $\Omega_\varphi=0.042$ which is in good agreement with the value shown in Fig.~\eqref{fig:EarlyDeFLRW}.
Of course, one should keep in mind that our approximations assumed that dark energy is subdominant in this regime, which is only true for $\mu\gtrsim1$. Existing upper bounds on EDE are model-dependent, but overall give $f_\text{EDE}\lesssim 10\%$ \cite{Poulin:2023lkg} at 95\%CL, with $f_\text{EDE}=\rho_\text{DE}/\rho_\text{total}$. In our case, this translates to $\mu\gtrsim 1.3$, which is consistent with the range of values allowed by our solutions.

Indeed, by studying the whole system via dynamical analysis~\cite{BigPaper:2023}, one can show that EDE regime corresponds to a matter dominated point in which the matter density scales as
\begin{equation}
\Omega_m = 1 -\frac{1}{6\mu^2} \,  ,
\end{equation}
in agreement with the semianalytical expression derived in Eq.~\eqref{Omega_varphi-approx}.
This is a saddle point for 
$\mu \leq -1/\sqrt{6}$  and $\lambda>6\mu$ or $\mu \geq 1/\sqrt{6}$ and $\lambda <6\mu$.
The sequence thus ends with a dark energy attractor satisfying
\begin{equation}
\Omega_{\varphi} = 1, \quad w_{\varphi} = \frac{\lambda^2}{3} -1\, ,
\end{equation}
that is also in concordance with the numerical results shown in  Fig.~\eqref{fig:EarlyDeFLRW}. This point produces an accelerated stage and is stable under the conditions
\begin{equation}\label{eq:lamb-mu-range}
- \sqrt{2} < \lambda < \sqrt{2} \,,\qquad  \mu  \lessgtr \frac{\lambda^2-2}{2\lambda} \, \text{ for } \lambda \lessgtr 0 \, ,
\end{equation}
which is automatically satisfied within our particular choice $\mu =2, \lambda=1$. The model also presents some interesting properties as oscillating equation of state for some particular range of parameters. A complete discussion of these properties is presented in~\cite{BigPaper:2023}.
 
\subsection{Spatial anisotropies}

\begin{figure}
\includegraphics[width=0.9\linewidth]{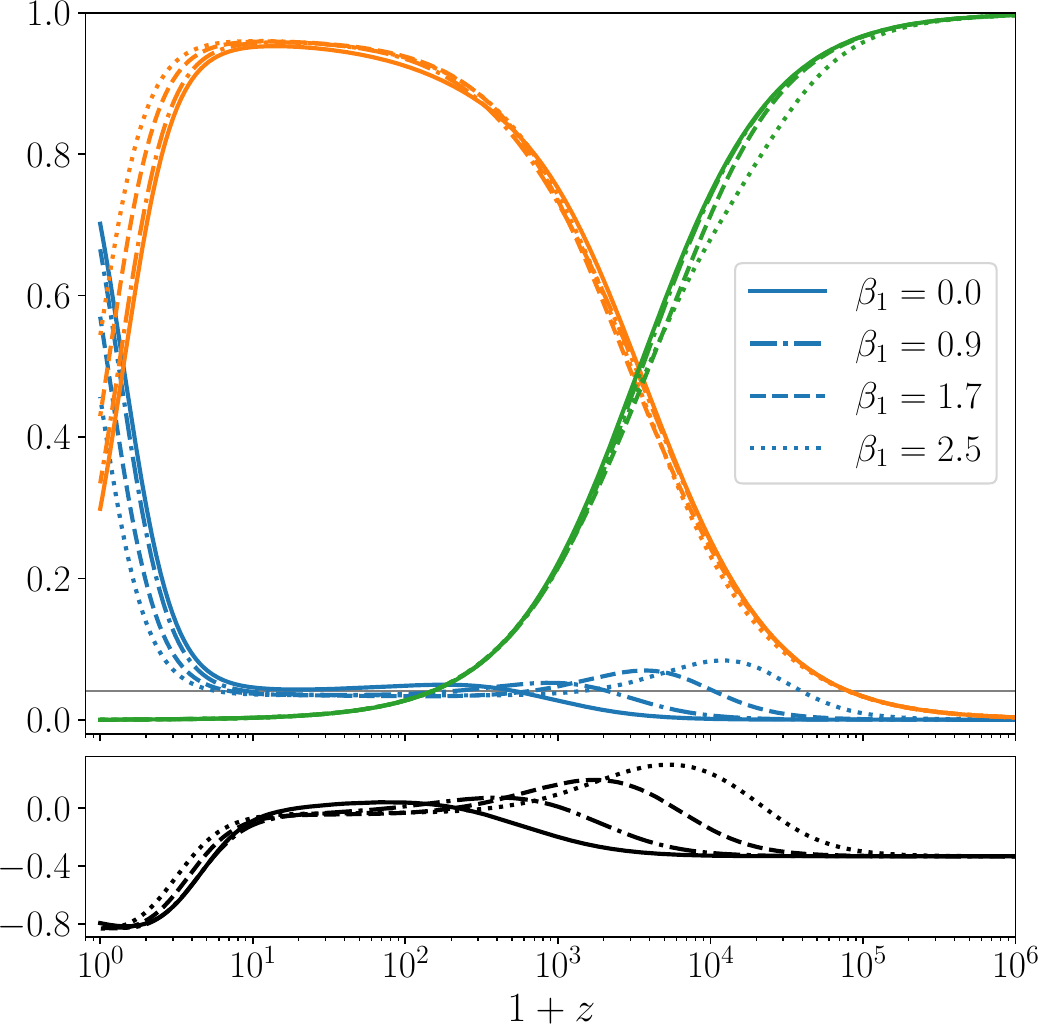}
\caption{Top: Background evolution of $\Omega_m$ (orange), $\Omega_r$ (green) and $\Omega_\varphi$ (blue) in a universe containing spatial anisotropies. Bottom: the equation of state of dark energy. We have adopted $\beta_1(0)=2\beta_2(0)$, and $\dot{\beta}_i(0)=0$ as initial conditions. The remaining parameters are the same used in the continuous lines of Fig.~\eqref{fig:EarlyDeFLRW}.}\label{fig:EarlyDeBI}
\end{figure}

As demonstrated, the gradients of the fields $\phi^I$ generally lead to the appearance of an anisotropic stress tensor, as described in \eqref{eq:pi-model}. In an expanding universe, small seeds of anisotropy will be sourced by this tensor, giving rise to increased anisotropies. These, in turn, amplify the gradients, which lead to greater anisotropic stresses in a self-feeding cycle. This is in fact the content of Eq.~\eqref{eq:trace-free}, suggesting that the late-time universe could exhibit significant anisotropies, potentially erasing the EDE phase found before. However, unlike everyday solids, cosmological solids are known to be quite insensitive to stretches and deformations~\cite{Bartolo:2013msa}, thus allowing any initial anisotropies to grow in a controlled way. Fig.~\eqref{fig:EarlyDeBI} shows the evolution of matter, radiation and dark energy in a universe with nonzero $\sigma^i_{\ph j}$ and $\pi^i_{\ph j}$. As we can see in this figure, the qualitative background behavior, including the estimate \eqref{Omega_varphi-approx}, is essentially the same as that of Fig.~\eqref{fig:EarlyDeFLRW}, where
$\sigma^i_{\ph j}=0=\pi^i_{\ph j}$. If anything, the presence of higher initial values of $\beta_i$ have the same effect as increasing the field-space metric amplitude $f_0$ in the isotropic case. This degenerate behavior follows from the product of
$\gamma f^2$ appearing in \eqref{def-rho_varphi}.

\begin{figure}
\includegraphics[width=0.9\linewidth]{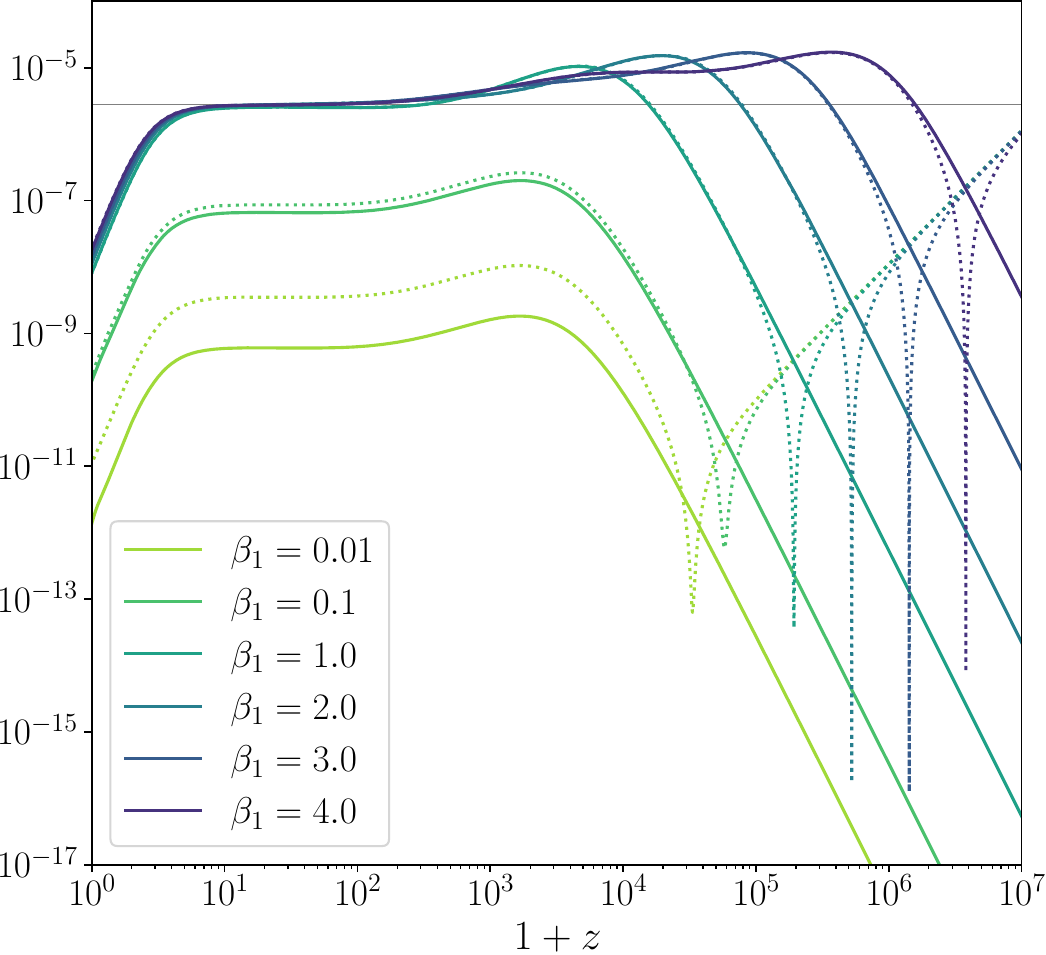}
\caption{Background evolution of the anisotropic fraction $\sigma^2/6H^2$ in a model with a curved solid. For a wide range of initial conditions on $\beta_i$ and $\dot{\beta}_i$, the late-time anisotropy has an upper bound depending on $\mu$. For this plot we have adopted $\beta_1 = 2\beta_2$, $\dot{\beta}_1=2\dot{\beta}_2=0$ (continuous lines),
$\dot{\beta}_1=2\dot{\beta}_2=10^{12}$ (dotted lines), and $\mu=10$. The remaining parameters are the same as in the continuous curves of Fig.~\eqref{fig:EarlyDeFLRW}. The horizontal thin line represents the upper bound given by the approximation \eqref{Shear-approx}.}\label{fig:Sigma2}
\end{figure}

\begin{figure}
\includegraphics[width=0.85\linewidth]{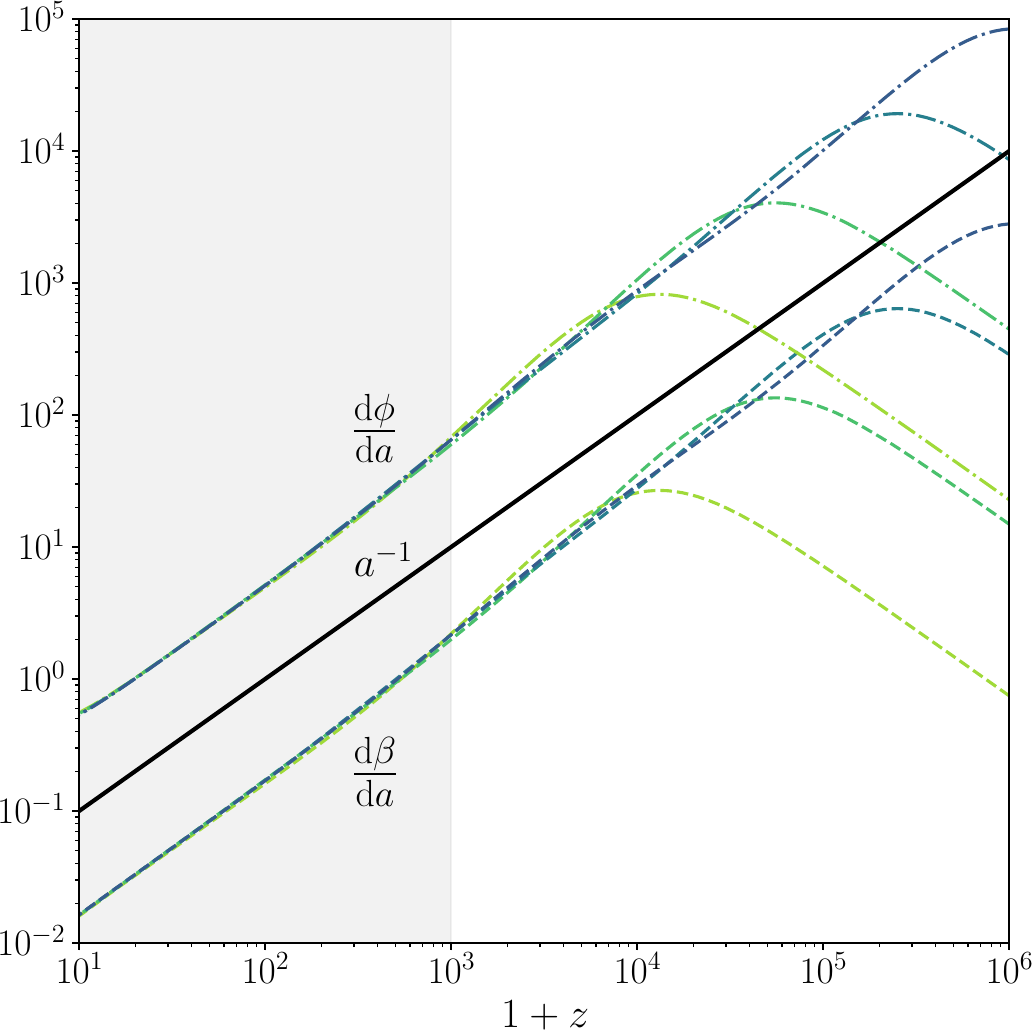}
\caption{Evolution of $\dd\phi/\dd a$ (dot dashed), $\dd\beta/\dd a$ (dashed), and $1/a$ (continuous) as a function of redshift, and for many initial values of $\beta_i$. During matter domination, all curves have approximately the same slope.}\label{fig:slope_a}
\end{figure}

Perhaps more surprising is the finding that, for a wide range of initial conditions on $\beta_i$ and $\dot{\beta}_i$, the fractional contribution of late-time anisotropy to the cosmic budget, given by $\sigma^2/6H^2$, can never exceed a limiting and nearly constant value. This feature is shown in Fig.~\eqref{fig:Sigma2} for several values of $\beta_i$ and $\dot{\beta}_i$. Similarly to $\Omega_\varphi$, this constant threshold is given by a simple function of $\mu$, and can also be estimated semianalytically during matter domination. For that, and with no loss of generality, we assume that $\beta_1 \gg 1$ at early times, so that $\gamma = \textrm{tr}(\gamma^{ij}) \gg \gamma^{11}$. During matter domination, and recalling \eqref{def-beta}, Eq. \eqref{eq:trace-free} can be approximated as ($i=1,2$)
\begin{equation}\label{eq:beta-approx}
 \frac{\dd ^2\beta_i}{\dd a^2} + \frac{5}{2a}\frac{\dd\beta_i}{\dd a} \approx -\frac{\gamma f_0^2e^{-2\mu\varphi}}{3H_0^2\Omega_m^0 a}\,.
\end{equation}
Note the similarity of this expression with \eqref{eq:varphi-approx}, which shows that the anisotropic scale factors $\beta_i$ behave as effective quintessence fields as long as $V$ is negligible. This is possible because the coupling between the field-space and spacetime metrics, which leads to a nonzero $\pi^1_{\ph 1}$, acts as an effective potential for $\beta_i$. Because of this, $\dot{\beta}_i^2$ tracks the evolution of $\Omega_m$, just like $\dot{\varphi}^2$ does. This means that $\dd\beta/\dd a = c/a$ for some constant $c$ during matter domination; see Fig.~\eqref{fig:slope_a}. Replacing this relation into the left hand side of \eqref{eq:beta-approx}, and the approximation \eqref{eq:approx_f} on the right, one immediately finds that $c=-1/6\mu^2$. Finally, using
$\dot{\beta}_1=\dot{\beta}_2 = -\dot{\beta}_3/2$, it follows that
\begin{equation}\label{Shear-approx}
\frac{\sigma^2}{6H^2} \lesssim \frac{1}{36\mu^4}\,,\qquad (10\lesssim z \lesssim 10^3)\,,
\end{equation}
where we have used an inequality to account for cases where $\beta_i\approx 0$ (see Fig.~\eqref{fig:Sigma2}.) Comparing this to~\eqref{Ricci} we see that $\sigma^2/6H^2\lesssim\ell_c^4$. In fact, we have found that
\begin{equation}
\frac{\sigma^2}{6H^2} \lesssim (\Omega_\varphi)^2\,.
\end{equation}
Thus, for $\mu\gtrsim1$, late-time anisotropies will be subdominant in a Bianchi I universe. Of course, these anisotropies can also be subdominant, regardless of $\mu$, if observations tell us that $\beta_i\approx 0$, in which case we would recover the FLRW results of the previous section. The take away lesson is that spatial anisotropies, \emph{if present}, will never exceed the dark energy fraction in these models.

\section{Linear perturbations}\label{sec:perturbations}
We have shown that the curved solid model is predictive at the background level, leading to a phase of early dark energy and (possibly) suppressed late-time anisotropies. The model is also predictive at the perturbative level, since the introduction of a solid changes the dispersion relation and the sound speed of perturbations, affecting the gravitational clustering of matter~\cite{Endlich:2012pz}. A thorough investigation of structure formation is complicated by the background shear~\eqref{def-beta}, which will couple scalar, vector and tensor modes, rendering the analysis much more cumbersome~\cite{Pitrou:2015iya}. We can nonetheless extract the model’s main footprints by considering the small scale limit of perturbations, where the cosmological effects of the expansion are negligible. Thus, in this section we adopt $\dot{\beta_i}=0=\dot{a}$; this is an acceptable approximation as long as we are considering Fourier modes deep inside the Hubble radius (see also \cite{Achucarro:2010da} for a related analysis in the context of an effective low energy description of multiple scalar fields with a hierarchical mass scale structure). In this Minkowskian regime, the field's perturbations obey (see Appendix~\ref{sec:perturbed-eq})
\begin{equation}\label{eq:perturbations}
\nabla^\mu\nabla_\mu\delta\phi^A + \mathcal{R}^A_{\ph CDB}\partial_\mu{\phi}^C\partial^\mu{\phi}^B\delta\phi^D - (\nabla^AV_C)\delta\phi^C = 0\,,
\end{equation}
where $\delta\phi^A$ represents the field's perturbation,
$\mathcal{R}^A{}_{BCD}$ is the field-space Riemann tensor, and $\nabla^\mu\nabla_\mu $ stands for $-\partial_{t}^2+\nabla^2$.

To simplify, we recycle our background convention and write $\delta\phi^0$ as $\delta\varphi$. We also introduce a vector notation where the three components $\delta\phi^I$ are written as $\delta\vec{\phi}$. Working in Fourier space and decomposing
$\delta\vec{\phi}$ into its longitudinal and transverse components,
\begin{equation}
 \delta\vec{\phi} = i\vec{k}\delta\phi_{L} + \delta\vec{\phi}_{T}\qquad (\vec{k}\cdot\delta\vec{\phi}_{T} = 0)\,,
\end{equation}
we find, after a straightforward computation, the following system of equations:
\begin{align}
 \delta\ddot{\varphi} + \left[k^2 + M^2_{\rm eff}\right]\delta\varphi & = -2\mu f^2k^2\delta\phi_{L}\,,\label{eq:delta-quint} \\
 \delta\ddot{\phi}_{L} + k^2\delta\phi_{L} -2\mu\dot{\varphi}\delta\dot{\phi}_{L} & = -2\mu \delta\varphi\,,\label{eq:delta-longit} \\
 \delta\ddot{\vec{\phi}}_{T} + k^2\delta\vec{\phi}_T - 2\mu\dot\varphi\delta\dot{\vec{\phi}}_T & = 0\,,\label{eq:delta-transv}
\end{align}
where we have defined $M_{\rm eff}^2=6\mu^2f^2+\lambda^2V$.

This system presents two new features following from the field-space curvature, and characterized by the parameter $\mu$: a coupling between the quintessence and longitudinal solid perturbations, $\delta\varphi$ and $\delta\phi_L$, and an intrinsic damping term in the longitudinal and transverse solid perturbations, $\delta\phi_L$ and $\delta\vec{\phi}_T$. These features are in effect crucial for the clustering properties of the model's perturbations, which can be unveiled by postulating oscillatory solutions and looking for the possible frequencies of oscillation.

Let us start with the scalar perturbations. Since $\delta\varphi$ and $\delta\phi_L$ are coupled, they share a common harmonic solution $e^{i\Omega t}$. The allowed frequencies are given as the roots of a fourth order polynomial in $\Omega$,
\begin{equation}\label{eq:polynomial}
 [\Omega^2-(k^2 + M^2_{\rm eff})][\Omega^2-k^2 + 2i\Omega\mu\dot\varphi] = 4\mu^2f^2k^2\,,
\end{equation}
with complex coefficients. Note the presence of cubic and linear terms in $\Omega$, which are generally absent in time-dependent multifield models~\cite{Achucarro:2010jv}. In the flat field-space limit ($\mu=0$), there are two independent frequencies corresponding to a light mode, $\Omega_{-}^2 = k^2$, and a heavy mode,
$\Omega_{+}^2=k^2+M_{\rm eff}^2$, both propagating with the speed of light. This is expected because when $\mu=0$ the model is linear and the perturbations evolve independently. At small scales, quintessence behave as a massive mode, with the mass being proportional to the second derivative of its background potential $V(\varphi)$. The solid perturbations, however, do not have a background potential (which, we recall, would violate spatial homogeneity) and thus evolve as light modes.

For arbitrary curvature (i.e., $\mu\neq0$), the solutions are generally complex, indicating the presence of instabilities in the scalar sector. Exact solutions to \eqref{eq:polynomial} can be obtained with algebraic packages, but the resulting expressions are too long to be useful. Nonetheless, it is important to note that the imaginary part of the frequencies result exclusively from the term $\mu\dot\varphi$, and not from the inhomogeneous term on the right-hand side of \eqref{eq:polynomial}. Indeed, if the former is absent, we find that the resulting frequencies,
\begin{equation}\label{eq:frequencies}
\Omega^2_{\pm}=\frac{1}{2}\left(2k^2+M_\text{eff}^2 \pm \sqrt{16\mu^2f^2k^2+M_\text{eff}^4}\right),
\end{equation}
are real since $2k^2+M_\text{eff}^2\geq \sqrt{16\mu^2f^2k^2+M^4_\text{eff}}$, as one can easily check using the definition of $M_\text{eff}^2$. In the presence of gravity, such small scale instabilities will feed scalar perturbations, and can lead to enhanced structures at large scales. Moreover, since this is a multif-field model, one might expect the appearance of isocurvature modes~\cite{Bassett:2005xm}, which could be another distinctive signature. 

The presence of instabilities is clearer in the case of transverse vector perturbations, which decouples from the scalar modes and thus oscillates with its own frequency. Inserting the ansatz $e^{i\omega t}$ in \eqref{eq:delta-transv}, the allowed frequencies,
\begin{equation}
\omega_{\pm} = -i\mu\dot\varphi\pm\sqrt{k^2-\mu^2\dot\varphi^2}\,,
\end{equation}
are clearly complex. At distances where gravitational effects can be neglected, we expect that $k^2\gg\mu^2\dot\varphi^2$. In this limit, $\omega_\pm$ simplifies to
\begin{equation}\label{eq:omega_approx}
\omega_{\pm} \approx -i\mu\dot\varphi \pm k\,.
\end{equation}
Therefore, small scale transverse perturbations propagate with the speed of light, and develop instabilities at distances of order $(\mu\dot\varphi)^{-1}$.

To estimate the typical scales involved, and to check whether this description is consistent with our numerical solutions, let us focus on low redshifts at which most structures are found, typically $z\lesssim3$. Taking ${\mu=2}$ as a reference, our numerical solutions give ${\dot\varphi\simeq 1.26 H_0}$ at $z=3$~\footnote{The value of $\dot\varphi$ varies little with the choice of $\mu$ in the range~\eqref{eq:lamb-mu-range} since the field solutions correspond to a stable point.}, which corresponds to $(\mu\dot\varphi)^{-1}\approx0.4H_0^{-1}$, roughly one half of the present Hubble radius. Adopting $H_0^{-1} = 3000h^{-1}\text{Mpc}$, this corresponds to a scale deep in the linear regime, ${\mu\dot\varphi~\sim10^{-3}h\text{Mpc}^{-1}}$, and justifies \emph{a posteriori} the small scale approximation leading to \eqref{eq:omega_approx}.

As is well known, primordial vector modes are not generated during inflation at first order in perturbations. This is because, differently from scalar and tensor modes, vector modes are not frozen after horizon crossing, but decay away. Moreover, even if they are produced by some mechanism at linear order, they do not act as sources of density perturbations, and thus are typically neglected in the standard analysis of structure formation. However, vectors modes can affect the number count of galaxies~\cite{Chen:2014bba,Durrer:2016jzq}, and their inclusion is also important to a proper data analysis of redshift-space distortion effects~\cite{Bonvin:2017req}. Here, we have presented a mechanism where vector modes generated from sub-horizon perturbations can be amplified at large scales. When gravitational perturbations are included, $\delta\vec{\phi}_T$, will source gravitational (transverse) vector modes, which can in principle be detectable with forthcoming surveys. We leave a thorough analysis of this important question for a future publication.

\section{Conclusion and Perspectives}\label{sec:conclusion}

We have presented a nonlinear multifield dark energy model where the fields evolve under
a nontrivial and maximally symmetric field-space metric. Differently from previous models in which the fields are either time or space-dependent, we have considered a model in which one field (quintessence) is time-dependent, while the others are given by pure spatial gradients, as in the effective field theory description of solids. The resulting background model is predictive, and can produce an early dark energy (EDE) component, as well as a suppressed fraction of late-time anisotropies --- both features being given by simple powers of the field-space curvature radius. We have shown that a percent level EDE component, such as the one employed in attempts to solve the $H_0$ tension, can be easily accomplished in the present scenario, and corresponds to a stable point in the space of possible solutions.

At the perturbative level, we have shown that the coupling between time and space-dependent fields also leads to interesting signatures, among which is the appearance of complex dispersion relations for the small scale perturbations. In particular, we have shown that small scale vector perturbations will develop tachyonic instabilities that can enhance structures at large scales, slightly below the present Hubble radius. Scalar perturbations will also become unstable, although the size of structures affected depend nontrivially on the scale of the perturbation. Incidentally, our findings corroborate earlier claims, made in the context of multifield inflation~\cite{Renaux-Petel:2015mga} and dark energy~\cite{Akrami:2020zfz}, that a sufficiently negative field-space curvature can lead to tachyonic instabilities, with important observational consequences. However, while in the context of time-dependent fields such instabilities arise from the effective mass term of fields evolving non-geodesically (in field-space), in the present context they result from natural couplings between time and space-dependent fields, and will always be present if the field-space is curved, regardless of the field-space trajectories. Overall, our findings represent an interesting observational window to test the curvature of the field-space.

\section*{Acknowledgments}
We thank Ben Normann and Mikjel Thorsrud for useful insights during the initial stages of this work. J.P.B.A., A.G. and C.A.V.T. acknowledge financial support from the Patrimonio Autónomo - Fondo Nacional de Financiamiento para la Ciencia, la Tecnología y la Innovación Francisco José de Caldas (MINCIENCIAS - COLOMBIA) Grant No. 110685269447 RC-80740-465- 2020, projects 69475, 69553 and 69723. T.S.P. thanks Conselho Nacional de Desenvolvimento Científico e Tecnológico (CNPq) and Coordenação de Aperfeiçoamento de Pessoal de Nível Superior (CAPES) for the financial support under grants 312869/2021-5 and 88881.709790/2022-01. J.P.B.A. and C.A.V.T. thank the Physics Department of Londrina State University for the hospitality during the development of this work.

\appendix

\section{Derivation of Eq.~\eqref{eq:perturbations}}\label{sec:perturbed-eq}
%

Suppose that $\phi^{A}_\lambda(x^\mu)$ is a one-parameter family of fields on a (flat) manifold 
$\mL{M}$, such that $\phi^A_0=\phi^A$ and $\left.\partial \phi^A_\lambda/\partial\lambda\right|_{\lambda=0}=\delta\phi^A$. The equation for the perturbations $\delta\phi^A$ can be defined as the first term in a Taylor expansion of Eq.~\eqref{eq:harmonic-eom} around $\lambda=0$. The computation can be simplified if we adopt Riemann normal coordinates in $\mL{N}$ such that, at $\lambda=0$, $G_{AB}=\delta_{AB}$ and $\partial_\lambda G_{AB}=0=\Gamma^A_{BC}$. Let us start with the kinetic term, which gives
\begin{align}\label{1st-term-perteq}
\left.\frac{\partial}{\partial\lambda}\nabla^\mu\nabla_\mu \phi^A_\lambda\right|_{\lambda=0} & \overset{!}{=}  \partial^\mu\partial_\mu\delta\phi^A + \partial_D\Gamma^{A}_{BC}\delta\phi^D\partial^\mu\phi^B\partial_\mu\phi^C\,, \nonumber \\ 
& \overset{!}{=} \nabla^\mu\nabla_\mu\delta\phi^A + \partial^\mu \Gamma^A_{DC}\delta\phi^C\partial_\mu\phi^D \nonumber \\ 
& \qquad+\mL{R}^A_{\;CDB}\partial_\mu\phi^C\partial^\mu\phi^B\delta\phi^D\,, 
\end{align}
where the exclamation point reminds us that we are working in a specific coordinate system. In the last equality we have added and subtracted $\partial\Gamma$ terms needed to introduce the Riemann tensor, and added $\Gamma=0$ terms needed to rewrite partial derivatives as covariant derivatives. The potential term can be computed in a similar fashion, giving:
\begin{align}\label{2st-term-perteq}
\left.\frac{\partial}{\partial\lambda} G^{AB} \partial_B V\right|_{\lambda=0} 
& \overset{!}{=} G^{AB}\partial_C(V_B)\delta\phi^C\,, \nonumber \\
& \overset{!}{=} G^{AB}(\nabla_B V^C)\delta\phi^C\,.
\end{align}
Expressions \eqref{1st-term-perteq} and \eqref{2st-term-perteq} are tensorial, and thus hold in any coordinate system. Their difference give \eqref{eq:perturbations}. Note that, if $V=0$, equation \eqref{eq:perturbations} is formally equivalent to the geodesic deviation equation (see also~\cite{Misner:1978am}).
\bibliographystyle{h-physrev4}
\bibliography{references}

\begin{thebibliography}{10}

\bibitem{Abdalla:2022yfr}
E.~Abdalla {\em et~al.},
\newblock JHEAp {\bf 34}, 49 (2022), [2203.06142].

\bibitem{Aluri:2022hzs}
P.~K. Aluri {\em et~al.},
\newblock Class. Quant. Grav. {\bf 40}, 094001 (2023), [2207.05765].

\bibitem{Tsujikawa:2013fta}
S.~Tsujikawa,
\newblock Class. Quant. Grav. {\bf 30}, 214003 (2013), [1304.1961].

\bibitem{Banerjee:2020xcn}
A.~Banerjee {\em et~al.},
\newblock Phys. Rev. D {\bf 103}, L081305 (2021), [2006.00244].

\bibitem{Dubovsky:2011sj}
S.~Dubovsky, L.~Hui, A.~Nicolis and D.~T. Son,
\newblock Phys. Rev. D {\bf 85}, 085029 (2012), [1107.0731].

\bibitem{Ballesteros:2016gwc}
G.~Ballesteros, D.~Comelli and L.~Pilo,
\newblock Phys. Rev. D {\bf 94}, 124023 (2016), [1603.02956].

\bibitem{Gruzinov:2004ty}
A.~Gruzinov,
\newblock Phys. Rev. D {\bf 70}, 063518 (2004), [astro-ph/0404548].

\bibitem{Endlich:2012pz}
S.~Endlich, A.~Nicolis and J.~Wang,
\newblock JCAP {\bf 10}, 011 (2013), [1210.0569].

\bibitem{Bartolo:2013msa}
N.~Bartolo, S.~Matarrese, M.~Peloso and A.~Ricciardone,
\newblock JCAP {\bf 08}, 022 (2013), [1306.4160].

\bibitem{Bucher:1998mh}
M.~Bucher and D.~N. Spergel,
\newblock Phys. Rev. D {\bf 60}, 043505 (1999), [astro-ph/9812022].

\bibitem{Verlinde:2016toy}
E.~P. Verlinde,
\newblock SciPost Phys. {\bf 2}, 016 (2017), [1611.02269].

\bibitem{Armendariz-Picon:2007umg}
C.~Armendariz-Picon,
\newblock JCAP {\bf 09}, 014 (2007), [0705.1167].

\bibitem{Motoa-Manzano:2020mwe}
J.~Motoa-Manzano, J.~Bayron Orjuela-Quintana, T.~S. Pereira and C.~A.
  Valenzuela-Toledo,
\newblock Phys. Dark Univ. {\bf 32}, 100806 (2021), [2012.09946].

\bibitem{Celoria:2017bbh}
M.~Celoria, D.~Comelli and L.~Pilo,
\newblock JCAP {\bf 09}, 036 (2017), [1704.00322].

\bibitem{Sasaki:1995aw}
M.~Sasaki and E.~D. Stewart,
\newblock Prog. Theor. Phys. {\bf 95}, 71 (1996), [astro-ph/9507001].

\bibitem{Langlois:2008mn}
D.~Langlois and S.~Renaux-Petel,
\newblock JCAP {\bf 04}, 017 (2008), [0801.1085].

\bibitem{Allen:2005ye}
L.~E. Allen, S.~Gupta and D.~Wands,
\newblock JCAP {\bf 01}, 006 (2006), [astro-ph/0509719].

\bibitem{Achucarro:2010jv}
A.~Achucarro, J.-O. Gong, S.~Hardeman, G.~A. Palma and S.~P. Patil,
\newblock Phys. Rev. D {\bf 84}, 043502 (2011), [1005.3848].

\bibitem{Achucarro:2010da}
A.~Achucarro, J.-O. Gong, S.~Hardeman, G.~A. Palma and S.~P. Patil,
\newblock JCAP {\bf 01}, 030 (2011), [1010.3693].

\bibitem{Kaiser:2012ak}
D.~I. Kaiser, E.~A. Mazenc and E.~I. Sfakianakis,
\newblock Phys. Rev. D {\bf 87}, 064004 (2013), [1210.7487].

\bibitem{Renaux-Petel:2015mga}
S.~Renaux-Petel and K.~Turzy\'nski,
\newblock Phys. Rev. Lett. {\bf 117}, 141301 (2016), [1510.01281].

\bibitem{Vardanyan:2015oha}
V.~Vardanyan and L.~Amendola,
\newblock Phys. Rev. D {\bf 92}, 024009 (2015), [1502.05922].

\bibitem{Akrami:2020zfz}
Y.~Akrami, M.~Sasaki, A.~R. Solomon and V.~Vardanyan,
\newblock Phys. Lett. B {\bf 819}, 136427 (2021), [2008.13660].

\bibitem{Cicoli:2020cfj}
M.~Cicoli, G.~Dibitetto and F.~G. Pedro,
\newblock Phys. Rev. D {\bf 101}, 103524 (2020), [2002.02695].

\bibitem{Anguelova:2021jxu}
L.~Anguelova, J.~Dumancic, R.~Gass and L.~C.~R. Wijewardhana,
\newblock JCAP {\bf 03}, 018 (2022), [2111.12136].

\bibitem{Eskilt:2022zky}
J.~R. Eskilt, Y.~Akrami, A.~R. Solomon and V.~Vardanyan,
\newblock Phys. Rev. D {\bf 106}, 023512 (2022), [2201.08841].

\bibitem{eells1964harmonic}
J.~Eells and J.~H. Sampson,
\newblock American journal of mathematics {\bf 86}, 109 (1964).

\bibitem{Misner:1978am}
C.~W. Misner,
\newblock Phys. Rev. D {\bf 18}, 4510 (1978).

\bibitem{eells1978report}
J.~Eells and L.~Lemaire,
\newblock Bulletin of the London mathematical society {\bf 10}, 1 (1978).

\bibitem{Pereira:2007yy}
T.~S. Pereira, C.~Pitrou and J.-P. Uzan,
\newblock JCAP {\bf 09}, 006 (2007), [0707.0736].

\bibitem{ellis2012relativistic}
G.~F. Ellis, R.~Maartens and M.~A. MacCallum,
\newblock {\em Relativistic cosmology} (Cambridge University Press, 2012).

\bibitem{martinez1995delta}
E.~Martinez-Gonzalez and J.~Sanz,
\newblock Astronomy and Astrophysics, v. 300, p. 346 {\bf 300}, 346 (1995).

\bibitem{Saadeh:2016sak}
D.~Saadeh, S.~M. Feeney, A.~Pontzen, H.~V. Peiris and J.~D. McEwen,
\newblock Phys. Rev. Lett. {\bf 117}, 131302 (2016), [1605.07178].

\bibitem{BigPaper:2023}
J.~P. Beltrán~Almeida, A.~Guarnizo, T.~S. Pereira and C.~A. Valenzuela-Toledo,
\newblock (2024).

\bibitem{Poulin:2023lkg}
V.~Poulin, T.~L. Smith and T.~Karwal,
\newblock Phys. Dark Univ. {\bf 42}, 101348 (2023), [2302.09032].

\bibitem{Pitrou:2015iya}
C.~Pitrou, T.~S. Pereira and J.-P. Uzan,
\newblock Phys. Rev. D {\bf 92}, 023501 (2015), [1503.01125].

\bibitem{Bassett:2005xm}
B.~A. Bassett, S.~Tsujikawa and D.~Wands,
\newblock Rev. Mod. Phys. {\bf 78}, 537 (2006), [astro-ph/0507632].

\bibitem{Chen:2014bba}
S.~Chen and D.~J. Schwarz,
\newblock Phys. Rev. D {\bf 91}, 043507 (2015), [1407.4682].

\bibitem{Durrer:2016jzq}
R.~Durrer and V.~Tansella,
\newblock JCAP {\bf 07}, 037 (2016), [1605.05974].

\bibitem{Bonvin:2017req}
C.~Bonvin, R.~Durrer, N.~Khosravi, M.~Kunz and I.~Sawicki,
\newblock JCAP {\bf 02}, 028 (2018), [1712.00052].

\end{thebibliography}

\end{document}